\documentclass[conference]{IEEEtran}
\IEEEoverridecommandlockouts

\usepackage{amsmath,graphicx,hyperref}
\usepackage{amssymb,amsfonts}
\usepackage{booktabs} 
\usepackage{subcaption}
\usepackage{multirow}
\usepackage{xcolor}
\usepackage{tikz}
\usepackage{pgfplots}
\pgfplotsset{compat=1.18}
\usetikzlibrary{positioning, arrows.meta, shapes.geometric, fit, decorations.pathreplacing, calc, backgrounds}

\usepackage{xcolor}

\def\BibTeX{{\rm B\kern-.05em{\sc i\kern-.025em b}\kern-.08em
    T\kern-.1667em\lower.7ex\hbox{E}\kern-.125emX}}

\title{SignDeepSC: A Semantic Signature-based Approach for Robust Semantic Communication
{\footnotesize}
\thanks{This work is supported by the American University of Beirut University Research Board and Vertically Integrated Projects Program.\par
\copyright~2026 IEEE.  Personal use of this material is permitted.  Permission from IEEE must be obtained for all other uses, in any current or future media, including reprinting/republishing this material for advertising or promotional purposes, creating new collective works, for resale or redistribution to servers or lists, or reuse of any copyrighted component of this work in other works.}
}
\author{Khalil Alhaj, Razane Tajeddine, and Hadi Sarieddeen\\
Department of Electrical and Computer Engineering\\
American University of Beirut, Beirut 1107 2020, Lebanon\\
kma90@mail.aub.edu, razane.tajeddine@aub.edu.lb, hadi.sarieddeen@aub.edu.lb
}

\begin{document}

%
\maketitle
\begin{abstract}
Semantic communication systems such as deep semantic communication (DeepSC) offer high efficiency but are vulnerable to adversarial attacks on their underlying neural networks. We address a physical-layer man-in-the-middle (MitM) threat in which an adversary injects perturbations into the transmitted signal to distort its meaning. We propose SignDeepSC, an architectural defense that achieves adversarial robustness without requiring explicit adversarial example generation during training. The approach is built on a perceiver-inspired semantic signature, a compact vector summary of the source features transmitted over a separate low-rate auxiliary channel. This signature is used by a self-repairing decoder that leverages cross-attention to correct distortions and can additionally drive a scrambler that shuffles the feature layout. We evaluate SignDeepSC over Rayleigh fading and additive white Gaussian noise channels under both single-step fast gradient sign method (FGSM) and iterative projected gradient descent (PGD) attacks. Under PGD ($\epsilon = 0.7$), at 12~dB signal-to-noise ratio with Rayleigh fading, SignDeepSC achieves a bilingual evaluation understudy (BLEU-4) score of 0.237 and bidirectional encoder representations from transformers (BERT) sentence similarity of 0.646, outperforming all baselines without degrading clean-channel performance, when the signature channel is well protected.

\end{abstract}

\section{Introduction}
\label{sec:intro}

\sloppy
Semantic communication aims to transmit the intended meaning of a message rather than its exact bit-level representation \cite{qin2022semantic, getu2024semantic}. Deep semantic communication (DeepSC) \cite{xie2021deep} is a representative framework that leverages a transformer-based encoder--decoder architecture \cite{vaswani2017attention} to jointly perform source and channel coding in an end-to-end manner, enabling efficient semantic transmission. However, its reliance on deep neural networks makes it inherently vulnerable to adversarial attacks \cite{goodfellow2014explaining, madry2017towards}. In practice, this vulnerability enables a physical-layer man-in-the-middle (MitM) attack, in which an adversary injects a perturbation into the transmitted signal, causing the receiver to output a message with altered meaning~\cite{sadeghi2019physical}.

Existing defenses broadly fall into two categories~\cite{guo2024survey}. Model-level defenses such as adversarial training \cite{goodfellow2014explaining, nan2023physical} and robust encoding~\cite{peng2024robust} modify the main transceiver and can degrade clean-channel accuracy. Auxiliary-channel defenses transmit side information to aid recovery. For example, the semantic protection (SemProtector) framework, and specifically its SemRECT module \cite{liu2023semprotector}, trains a Wasserstein generative adversarial network (GAN) on clean encoder outputs and requires computationally costly per-sentence gradient optimization at inference. To the best of our knowledge, the impact of side-channel quality on semantic communication robustness under auxiliary-channel defenses has not been studied.

To address these gaps, we propose SignDeepSC, an architecture-level defense for the physical-layer MitM threat that achieves robustness without requiring adversarial noise injection during training. Instead of modifying the transceiver or running costly per-sentence optimization, SignDeepSC transmits a compact perceiver-inspired \cite{jaegle2021perceiver} semantic signature over a side channel and feeds it into a self-repairing decoder that introduces signature-conditioned cross-attention at each transformer decoder layer. The use of clean auxiliary information as cross-attention keys and values to condition a corrupted representation follows the same principle as in latent diffusion models~\cite{rombach2022high} and reference-based restoration~\cite{yang2020learning}: because the signature is transmitted over an independent channel, it is less affected by attacks on the main signal; each decoder layer then queries this clean reference via cross-attention to correct corrupted features before producing its output. Inspired by \cite{chen2025shuffling}, we also include a scrambler that derives a permutation seed $\kappa$ from the signature to shuffle the feature layout, disrupting the spatial locality exploited by gradient-based attacks. SignDeepSC preserves clean-channel accuracy and requires only a single forward pass at inference. We additionally present a sensitivity analysis of side-channel quality on defense performance.

Our main contributions are as follows:
\begin{itemize}\setlength{\itemsep}{0pt}\setlength{\parsep}{0pt}
    \item A perceiver-inspired semantic signature generator for compact side-channel transmission.
    \item A self-repairing decoder with signature-conditioned cross-attention and gated fusion, and a scrambler that obfuscates feature layout via a signature-derived invariant~$\kappa$.
    \item Evaluation across Rayleigh and additive white Gaussian noise (AWGN) channels, including a side-channel sensitivity study for auxiliary-channel defenses, a component ablation, and a joint attack analysis considering white-box attacks on both channels.
\end{itemize}

\section{System Model and Problem Formulation}
\label{sec:system}

We consider the semantic communication system shown in Fig.~\ref{fig:architecture}. Source text $\mathbf{x}$ of length $L$ is encoded into semantic features $\mathbf{Z} = f_{\text{enc}}(\mathbf{x})$, scrambled into $\mathbf{Z}_{\text{sc}}$, and transmitted as $\mathbf{T}_{\text{x}} = g_{\text{enc}}(\mathbf{Z}_{\text{sc}})$ over a Rayleigh fading channel $h \sim \mathcal{CN}(0,1)$ with additive white Gaussian noise. The receiver recovers $\mathbf{Z}' = g_{\text{dec}}(\mathbf{R}_{\text{x}})$ and reconstructs the output $\hat{\mathbf{x}} = f_{\text{dec}}(\mathbf{Z}', \mathbf{s}')$, where $\mathbf{s}'$ is the received signature transmitted over the independent side channel.

We investigate a physical-layer MitM attack under the following threat model:
\begin{itemize}\setlength{\itemsep}{0pt}\setlength{\parsep}{0pt}
    \item \textit{Knowledge:} The adversary has white-box access only to the architecture and weights of the shared main-channel path; it cannot access the signature generator, side-channel codec, or scrambler/inverse-scrambler, since these depend on content carried by the inaccessible side channel, and hence cannot include them in its gradient computation.
    \item \textit{Access:} The adversary can intercept and modify the main data channel but \emph{cannot} access the signature side-channel. This models a physically separated link (e.g., a different frequency band or a low-rate coded sub-carrier) for the low-bandwidth signature. We relax this assumption in Section~\ref{sec:sig_attack} by evaluating joint attacks on both channels.
\end{itemize}
Under this threat model, the received signal becomes $\mathbf{R}_{\text{x}} = h(\mathbf{T}_{\text{x}} + \boldsymbol{\delta})$, and the receiver, unaware of the attack, produces an output $\hat{\mathbf{x}}'$ with distorted meaning.

\section{The Proposed SignDeepSC Approach}
\label{sec:approach}

\begin{figure*}[htb]
 \centering
 \resizebox{\textwidth}{!}{\begin{tikzpicture}[
  node distance=1.1cm and 1.1cm,
  font=\Large\sffamily,
  proc/.style  = {rectangle, rounded corners=5pt,
                  fill=blue!7, draw=blue!32!gray, line width=0.7pt,
                  align=center, text width=2.5cm, minimum height=1.6cm,
                  inner sep=5pt},
  scrblk/.style= {rectangle, rounded corners=5pt,
                  fill=green!7, draw=green!38!gray, line width=0.7pt,
                  align=center, text width=3.0cm, minimum height=1.6cm,
                  inner sep=5pt},
  chblk/.style = {ellipse, dashed,
                  fill=orange!6, draw=orange!42!gray, line width=0.6pt,
                  minimum width=2.5cm, minimum height=2.6cm,
                  align=center, inner sep=3pt},
  advblk/.style= {rectangle, rounded corners=5pt, densely dashed,
                  fill=red!5, draw=red!42!gray, line width=0.7pt,
                  align=center, text width=2.4cm, minimum height=1.1cm,
                  inner sep=4pt},
  arr/.style   = {-{Latex[length=9pt, width=7pt]},
                  line width=1.1pt, gray!65!black},
  darr/.style  = {-{Latex[length=8pt, width=6pt]},
                  line width=1.0pt, densely dashed, gray!50!black},
  advarr/.style= {-{Latex[length=11pt, width=8pt]},
                  line width=1.4pt, red!55!black},
  lbl/.style   = {font=\Large, text=gray!65!black, inner sep=1pt},
  io/.style    = {font=\LARGE\sffamily\itshape, text=gray!60!black},
  grp/.style   = {draw=gray!32, dashed, rounded corners=8pt,
                  inner sep=10pt, fill=gray!3, line width=0.5pt},
  grplbl/.style= {font=\Large\sffamily\bfseries, text=gray!50},
]

\node[proc]                          (semenc)  {Semantic\\encoder};
\node[scrblk, right=of semenc]       (scr)     {Scrambler};
\node[proc,   right=of scr]          (chenc)   {Channel\\encoder};
\node[chblk,  right=1.4cm of chenc]  (chan)    {Wireless\\channel};
\node[proc,   right=1.4cm of chan]   (chdec)   {Channel\\decoder};
\node[scrblk, right=of chdec]        (iscr)    {Scrambler$^{-1}$};

\node[proc, below right=1.4cm and 1.1cm of iscr] (semdec) {Self-repairing\\decoder};

\node[proc,  below=2.8cm of semenc]  (siggen)   {Signature\\generator};
\node[proc,  below=2.8cm of chenc]   (sigchenc) {Channel\\encoder};
\node[chblk] (sigchan)  at (chan  |- sigchenc)   {Sig.\\channel};
\node[proc,  below=2.8cm of chdec]   (sigchdec) {Channel\\decoder};

\node[advblk, above=1.0cm of chan]   (adv) {Adversary};

\node[io, left=0.6cm of semenc]      (src) {source};
\node[io, right=0.5cm of semdec]    (dst) {output};

\draw[advarr] (adv.south) --
    node[right, lbl, text=red!55!black, pos=0.45]{$\boldsymbol{\delta}$}
    (chan.north);

\draw[arr] (src)    -- (semenc);
\draw[arr] (semenc) -- node[above,lbl]{$\mathbf{Z}$}               (scr);
\draw[arr] (scr)    -- node[above,lbl]{$\mathbf{Z}_{\mathrm{sc}}$} (chenc);
\draw[arr] (chenc)  -- node[above,lbl]{$\mathbf{T}_{\!\mathbf{X}}$}(chan);
\draw[arr] (chan)   -- node[above,lbl]{$\mathbf{R}_{\!\mathbf{X}}$}(chdec);
\draw[arr] (chdec)  -- node[above,lbl]{$\mathbf{Z}'_{\mathrm{sc}}$}(iscr);

\draw[arr] (iscr.east) -- node[above,lbl]{$\mathbf{Z}'$}
    ++(0.5,0) |- (semdec.west);

\draw[arr] (semdec) -- (dst);

\draw[arr] (semenc.south) -- (siggen.north);
\draw[arr] (siggen)   -- node[above,lbl]{$\mathbf{s}_{\mathrm{clean}}$}
                          (sigchenc);
\draw[arr] (sigchenc) -- node[above,lbl]{$\mathbf{t}_{\mathrm{sig}}$}
                          (sigchan);
\draw[arr] (sigchan)  -- node[above,lbl]{$\mathbf{r}_{\mathrm{sig}}$}
                          (sigchdec);

\draw[arr] (sigchdec.east) -|
    node[pos=0.25, above, lbl]{$\mathbf{s}'$} (semdec.south);

\draw[darr] (siggen.north east) --
    node[pos=0.40, right, lbl]{$\mathbf{s}_{\mathrm{clean}}$} (scr.south);
\draw[darr] (sigchdec.north) --
    node[pos=0.44, right, lbl]{$\mathbf{s}'$}                 (iscr.south);

\begin{pgfonlayer}{background}
  \node[grp, fit=(semenc)(scr)(chenc)(siggen)(sigchenc),
        label={[grplbl, anchor=north west, xshift=5pt, yshift=-3pt]
               north west:TRANSMITTER}] {};
  \node[grp, fit=(chdec)(iscr)(semdec)(sigchdec),
        label={[grplbl, anchor=north west, xshift=5pt, yshift=-3pt]
               north west:RECEIVER}] {};
\end{pgfonlayer}

\end{tikzpicture}}
     \caption{Proposed SignDeepSC architecture. Source text is encoded into $\mathbf{Z}$, permuted by a signature-keyed scrambler into $\mathbf{Z}_{\mathrm{sc}}$, and transmitted as $\mathbf{T}_{\mathbf{X}}$. In parallel, $\mathbf{Z}$ is compressed into $\mathbf{s}_{\mathrm{clean}}$, encoded, and sent as $\mathbf{t}_{\mathrm{sig}}$ over a side channel. At the receiver, $\mathbf{Z}'$ is recovered from $\mathbf{Z}_{\mathrm{sc}}$ using $\mathbf{s}'$, and the self-repairing decoder reconstructs the output from $\mathbf{Z}'$ and $\mathbf{s}'$. An adversary injects $\boldsymbol{\delta}$ into the main channel, while the side channel remains inaccessible.}
 \label{fig:architecture}
\end{figure*}
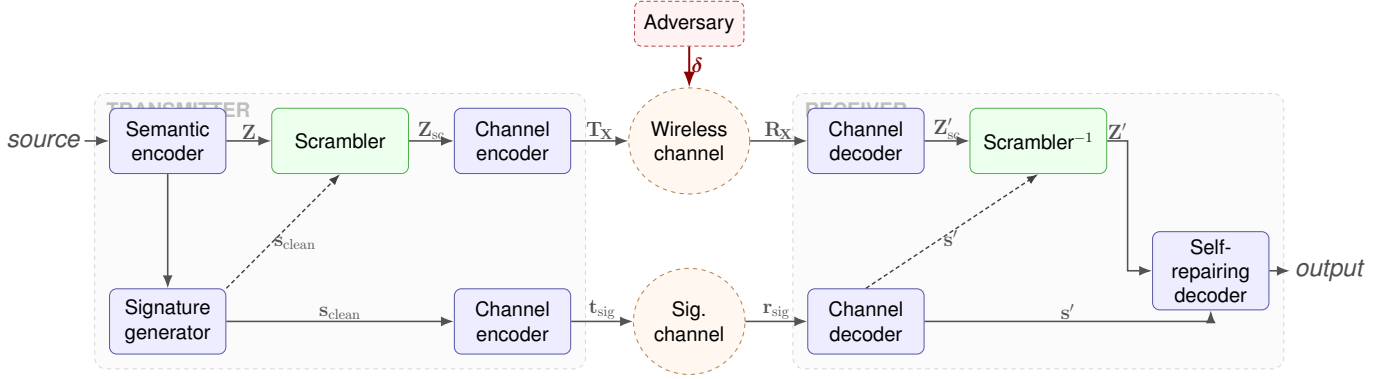

As shown in Fig.~\ref{fig:architecture}, SignDeepSC adds three components to the baseline DeepSC transceiver: a signature generator, a signature-conditioned scrambler, and a self-repairing decoder. The core insight is that adversarial attacks corrupt a high-dimensional feature space, while the compact signature provides a low-dimensional clean anchor; cross-attention then projects corrupted features toward the clean manifold defined by this anchor, effectively converting robustness into representation alignment.

\subsection{Signature Generator}
The signature generator maps the semantic feature matrix $\mathbf{Z} \in \mathbb{R}^{L \times d_{\text{model}}}$ to a compact signature $\mathbf{s}_{\text{clean}} \in \mathbb{R}^{d_{\text{sig}}}$. Inspired by the perceiver \cite{jaegle2021perceiver}, a learnable aggregator token $\mathbf{p}_{\text{agg}} \in \mathbb{R}^{d_{\text{model}}}$ is prepended to form $\mathbf{Z}_{\text{aug}} = [\mathbf{p}_{\text{agg}}, \mathbf{Z}]$ and processed via multi-head self-attention \cite{vaswani2017attention}. The updated token $\mathbf{p}'_{\text{agg}}$, which attends to all positions, is then projected as
\begin{equation}
\mathbf{s}_{\text{clean}} = \mathbf{W}_{\text{proj}} \mathbf{p}'_{\text{agg}} + \mathbf{b}_{\text{proj}},
\end{equation}
where $\mathbf{W}_{\text{proj}} \in \mathbb{R}^{d_{\text{sig}} \times d_{\text{model}}}$. The resulting $\mathbf{s}_{\text{clean}}$ is passed through a dedicated channel encoder for transmission.

\subsection{Signature-Conditioned Scrambler}
\label{sec:scrambler}
To defend against attacks that exploit feature arrangement, we permute the feature sequence using a signature-derived invariant $\kappa$. We define $\kappa$ as a \emph{robust signature invariant} computed via a difference hash (dHash) of $\mathbf{s} \in \mathbb{R}^{d_{\text{sig}}}$:
\begin{equation}
\kappa \;=\; \sum_{i=1}^{N_{\text{bits}}}
\mathbf{1}\!\Big( s_{i+\Delta} - s_i > 0 \Big) \cdot 2^{\,i-1}.
\end{equation}
Because $\kappa$ encodes only relative sign differences, it remains stable under moderate channel noise. Both the transmitter and receiver independently compute the same $\kappa$ and apply inverse permutations. The hash-based permutation acts as a heuristic defense that disrupts gradient-based attacks by breaking feature locality; as this relies on a non-differentiable operation, it does not provide cryptographic security guarantees and may raise concerns of gradient masking~\cite{athalye2018obfuscated}, though it remains robust to signature noise (Fig.~\ref{fig:sensitivity}) and requires no additional training or channel.

\subsection{The Self-Repairing Decoder}

The key idea is to provide the decoder with clean reference information at every layer. The signature tokens act as uncorrupted keys and values that each decoder position can attend to. A dedicated multi-head cross-attention block is inserted after each self-attention layer. We note that using $\mathbf{s}'$ directly as a single key collapses the softmax to unity, resulting in vanishing key and value gradients. To restore attention diversity, a \emph{signature unpacker} at the receiver splits $\mathbf{s}'$ into $N$ chunks $\mathbf{s}'_i \in \mathbb{R}^{c}$, where $c = d_{\text{sig}}/N$ and $i = 1,\ldots,N$, and projects each:
\begin{align}
\mathbf{t}_i &= \mathrm{LN}(\mathbf{W}_{\text{up}}\,\mathbf{s}'_i + \mathbf{b}_{\text{up}}), \\
\mathbf{S} &= [\mathbf{t}_1;\ldots;\mathbf{t}_N] \in \mathbb{R}^{N \times d_{\text{model}}},
\end{align}
where $\mathrm{LN}(\cdot)$ denotes LayerNorm (per-token zero-mean, unit-variance normalization) and $\mathbf{W}_{\text{up}} \in \mathbb{R}^{d_{\text{model}} \times c}$, with no additional transmission cost. The decoder features $\mathbf{X}_{\text{dec}}$ query the $N$ tokens:
\begin{align}
\mathbf{Q} &= \mathbf{X}_{\text{dec}}\mathbf{W}_Q,\quad \mathbf{K}=\mathbf{S}\mathbf{W}_K,\quad \mathbf{V}=\mathbf{S}\mathbf{W}_V, \\
\mathbf{A}_{\text{sig}} &= \mathrm{softmax}\!\left( \frac{\mathbf{Q}\mathbf{K}^\top}{\sqrt{d_k}} \right)\mathbf{V}.
\end{align}
The correction is fused via a gated recurrent unit (GRU)-style gate~\cite{cho2014learning}:
\begin{align}
\mathbf{g} &= \sigma(\mathbf{W}_g [\mathbf{X}_{\text{dec}} \| \mathbf{A}_{\text{sig}}] + \mathbf{b}_g), \label{eq:gate}\\
\mathbf{X}'_{\text{dec}} &= \mathrm{LN}\!\left((1{-}\mathbf{g}) \odot \mathbf{X}_{\text{dec}} + \mathbf{g} \odot \mathbf{A}_{\text{sig}}\right), \label{eq:repair}
\end{align}
where $\sigma$ denotes the sigmoid function, $\|$ concatenation, and $\odot$ element-wise multiplication. The bias $\mathbf{b}_g$ is positively initialized to favor reliance on signature. This block is placed before encoder cross-attention, priming the decoder with clean signal information before processing the corrupted features.

\subsection{Channel Transmission and Training}

We use a single signature vector with $d_{\text{sig}} = d_{\text{model}}$, providing a compact summary of the $L \times d_{\text{model}}$ feature matrix. Both $\mathbf{Z}$ and $\mathbf{s}_{\text{clean}}$ are passed through dedicated channel encoders for compression and transmission. We freeze the original DeepSC sender weights and train only the signature generator and self-repairing decoder, ensuring backward compatibility. Training follows a two-phase strategy: Phase~1 optimizes the signature channel codec alone under a mean squared error (MSE) loss to maximize dHash stability, achieving $>$99\% hash match at a signature signal-to-noise ratio (SNR) of $\text{SNR}_{\text{sig}}{=}18$\,dB; Phase~2 freezes the Phase-1 codec and fine-tunes all remaining components jointly.

The signature is transmitted over a secure side-channel. To encourage the decoder to rely on the signature, we employ a \emph{memory replacement} strategy: during training, the main input $\mathbf{Z}'$ is stochastically replaced with Gaussian noise $\mathbf{N}$ using $m \sim \text{Bernoulli}(p_{\text{drop}})$,
\begin{equation}
\tilde{\mathbf{Z}}' = (1 - m) \cdot \mathbf{Z}' + m \cdot \mathbf{N},
\end{equation}
causing $p_{\text{drop}}$ to be annealed during training, and encouraging the decoder to use $\mathbf{s}'$ for reconstruction.

\section{Experimental Setup}
\label{sec:setup}

\subsection{Adversarial Attack Models}

Both attacks construct an additive perturbation $\boldsymbol{\delta}$ injected into the transmitted signal $\mathbf{T}_{\text{x}}$, constrained by $\|\boldsymbol{\delta}\|_\infty \leq \epsilon$, that maximizes the end-to-end semantic loss $\mathcal{L}_{\text{sem}}(\mathbf{T}_{\text{x}}+\boldsymbol{\delta})$.

FGSM \cite{goodfellow2014explaining} computes the perturbation in a single gradient step:
\begin{equation}
\boldsymbol{\delta}_{\text{FGSM}} = \epsilon \cdot \mathrm{sign}\!\left(\nabla_{\mathbf{T}_{\text{x}}} \mathcal{L}_{\text{sem}}\right). \label{eq:fgsm}
\end{equation}

PGD \cite{madry2017towards} refines the perturbation over $T$ projected steps, initialized from $\boldsymbol{\delta}_{0} \sim \mathcal{U}[-\epsilon, \epsilon]$:
\begin{equation}
\boldsymbol{\delta}_{t+1} = \Pi_{\epsilon}\!\left(\boldsymbol{\delta}_{t} + \alpha\cdot\mathrm{sign}\!\left(\nabla_{\mathbf{T}_{\text{x}}} \mathcal{L}_{\text{sem}}(\mathbf{T}_{\text{x}}+\boldsymbol{\delta}_{t})\right)\right), \label{eq:pgd}
\end{equation}
where $\Pi_{\epsilon}$ projects onto the $\ell_\infty$-ball of radius $\epsilon$, $\alpha = \epsilon/T$ is the step size, and $T{=}10$.

\subsection{DeepSC Architecture}

All models share the same base DeepSC architecture: a 4-layer transformer with $d_{\text{model}}{=}128$ and 8 attention heads, trained on Europarl-v7 English text (maximum sequence length 30). Training uses Rayleigh fading channels; we additionally evaluate on AWGN. To ensure a fair comparison, \emph{neither SignDeepSC nor SemRECT is trained with adversarial perturbations}; adversarial training (AdvTrain) uses PGD-10 at $\epsilon{=}0.1$ during training, following standard mixed adversarial training~\cite{nan2023physical}. All models are then evaluated at the stronger $\epsilon{=}0.7$ to assess out-of-distribution robustness. Each experiment is repeated over three independent Monte Carlo seeds; results report mean $\pm$ standard deviation. The signature is transmitted at $\text{SNR}_{\text{sig}}{=}20$\,dB (trained at 18\,dB). Channel encoders compress to $d_c{=}16$, the signature unpacker uses $N{=}8$ tokens, $b_g{=}2.0$, dHash stride $\Delta{=}4$ with $N_{\text{bits}}{=}16$, and $p_{\text{drop}}$ is annealed from 0.7 to 0.1.

\subsection{Baselines}
\label{sec:baselines}
We compare SignDeepSC against three baselines:
\begin{itemize}\setlength{\itemsep}{0pt}\setlength{\parsep}{0pt}
    \item DeepSC \cite{xie2021deep}: The undefended base model.
    \item AdvTrain \cite{goodfellow2014explaining, nan2023physical}: The base DeepSC model fine-tuned with mixed adversarial training, $\mathcal{L} \!=\! (1{-}\alpha)\mathcal{L}_{\text{clean}} \!+\! \alpha\mathcal{L}_{\text{adv}}$, with $\alpha{=}0.5$ and 10-step PGD perturbations at $\epsilon{=}0.1$.
    \item SemRECT \cite{liu2023semprotector}: A GAN-based calibration defense in which the receiver fuses the GAN reconstruction with corrupted features via a simple calibrator MLP.
\end{itemize}
SemRECT solves, per sentence:
\begin{equation}
\mathbf{z}^* = \underset{\mathbf{z}}{\arg\min}\;\mathcal{L}_{\text{sem}}(G(\mathbf{z})) + \lambda\,\mathcal{L}_{\text{cal}}(\mathbf{z}),
\end{equation}
where $G$ is the GAN generator and $\mathcal{L}_{\text{cal}}$ is the calibrator loss. Our SemRECT implementation follows~\cite{liu2023semprotector}, with hyperparameters tuned for our dataset, and assumes a simple MLP calibrator, as the original work does not specify the exact architecture.

\subsection{Overhead Analysis}
SignDeepSC adds three lightweight components to the base DeepSC transceiver: (i)~a two-layer signature generator, (ii)~a dedicated signature channel codec, and (iii)~one cross-attention block with gated fusion per decoder layer, totaling approximately 967K trainable parameters (9\% of the 10.58M base). After channel encoding ($d_c$ dimensions), the main data occupies $L \times d_c$ channel uses, and the signature adds only $d_c$, corresponding to a $3.3\%$ bandwidth overhead for the typical $L{=}30$ setting. The signature summarizes $L \times d_{\text{model}}$ features into a single $d_{\text{sig}}$-dimensional vector, while cross-attention selectively repairs corrupted dimensions.
In contrast to SemRECT, which reconstructs corrupted features via a GAN-scale generator requiring per-sentence gradient optimization, SignDeepSC replaces GAN-based reconstruction with attention conditioning~\cite{rombach2022high} at a single forward pass.

\section{Results and Analysis}
\label{sec:results}

We evaluate performance using the bilingual evaluation understudy (BLEU-4) score for grammatical accuracy and a bidirectional encoder representations from transformers (BERT)-based sentence similarity (Sim) metric for semantic fidelity. Unless otherwise noted, adversarial evaluations use $\epsilon=0.7$ over a Rayleigh fading channel.

\begin{figure*}[t]
\centering
\pgfplotsset{
    every axis/.append style={
        width=0.48\textwidth, height=4.4cm,
        grid=major, grid style={gray!30},
        xlabel={SNR (dB)}, xtick={0,3,6,9,12,15,18},
        tick label style={font=\scriptsize},
        label style={font=\small},
    },
    deepsc/.style={gray, mark=*, mark size=2pt, very thick},
    advtrain/.style={orange!80!black, mark=square*, mark size=2pt, very thick},
    semrect/.style={violet, mark=triangle*, mark size=2.5pt, very thick},
    signdeepsc/.style={blue!70!black, mark=diamond*, mark size=2.5pt, very thick},
}
\begin{tabular}{@{}cc@{}}
\begin{tikzpicture}
\begin{axis}[ylabel={BLEU-4}, title={\small Clean -- BLEU-4}, ymin=0, ymax=0.95]
\addplot[deepsc] coordinates {(0,0.2244)(3,0.3387)(6,0.5235)(9,0.5758)(12,0.6691)(15,0.7179)(18,0.7388)};
\addplot[advtrain] coordinates {(0,0.1846)(3,0.2701)(6,0.3425)(9,0.4212)(12,0.5279)(15,0.5278)(18,0.5612)};
\addplot[semrect] coordinates {(0,0.2754)(3,0.4518)(6,0.5147)(9,0.6273)(12,0.7016)(15,0.7288)(18,0.7589)};
\addplot[signdeepsc] coordinates {(0,0.3890)(3,0.5517)(6,0.6364)(9,0.7662)(12,0.7630)(15,0.8286)(18,0.8262)};
\end{axis}
\end{tikzpicture}
&
\begin{tikzpicture}
\begin{axis}[ylabel={Similarity}, title={\small Clean -- Similarity}, ymin=0.55, ymax=0.95]
\addplot[deepsc] coordinates {(0,0.5962)(3,0.6614)(6,0.7638)(9,0.7864)(12,0.8383)(15,0.8646)(18,0.8752)};
\addplot[advtrain] coordinates {(0,0.6002)(3,0.6601)(6,0.6969)(9,0.7347)(12,0.7923)(15,0.7924)(18,0.8093)};
\addplot[semrect] coordinates {(0,0.5983)(3,0.7056)(6,0.7407)(9,0.7950)(12,0.8378)(15,0.8515)(18,0.8660)};
\addplot[signdeepsc] coordinates {(0,0.7114)(3,0.7909)(6,0.8277)(9,0.8881)(12,0.8860)(15,0.9165)(18,0.9160)};
\end{axis}
\end{tikzpicture}
\\[2pt]
\begin{tikzpicture}
\begin{axis}[ylabel={BLEU-4}, title={\small Under Attack -- BLEU-4}, ymin=0, ymax=0.35]
\addplot[deepsc] coordinates {(0,0.0136)(3,0.0295)(6,0.0372)(9,0.0528)(12,0.0613)(15,0.0613)(18,0.0703)};
\addplot[advtrain] coordinates {(0,0.0396)(3,0.0510)(6,0.0822)(9,0.1026)(12,0.1055)(15,0.1085)(18,0.1288)};
\addplot[semrect] coordinates {(0,0.0390)(3,0.0758)(6,0.1071)(9,0.1425)(12,0.1864)(15,0.2015)(18,0.2019)};
\addplot[signdeepsc] coordinates {(0,0.0885)(3,0.1370)(6,0.1794)(9,0.2113)(12,0.2365)(15,0.2590)(18,0.2750)};
\addplot[gray, dashed, very thick, mark=o, mark size=2pt, every mark/.append style={solid}] coordinates {(0,0.0072)(3,0.0114)(6,0.0165)(9,0.0236)(12,0.0281)(15,0.0331)(18,0.0331)};
\addplot[orange!80!black, dashed, very thick, mark=square, mark size=2pt, every mark/.append style={solid}] coordinates {(0,0.0127)(3,0.0212)(6,0.0243)(9,0.0278)(12,0.0309)(15,0.0337)(18,0.0353)};
\addplot[violet, dashed, very thick, mark=triangle, mark size=2.5pt, every mark/.append style={solid}] coordinates {(0,0.0200)(3,0.0413)(6,0.0656)(9,0.0981)(12,0.1234)(15,0.1414)(18,0.1437)};
\addplot[blue!70!black, dashed, very thick, mark=diamond, mark size=2.5pt, every mark/.append style={solid}] coordinates {(0,0.0763)(3,0.1244)(6,0.1690)(9,0.1891)(12,0.2511)(15,0.2606)(18,0.2809)};
\end{axis}
\end{tikzpicture}
&
\begin{tikzpicture}
\begin{axis}[ylabel={Similarity}, title={\small Under Attack -- Similarity}, ymin=0.4, ymax=0.7]
\addplot[deepsc] coordinates {(0,0.4648)(3,0.4913)(6,0.4999)(9,0.5207)(12,0.5291)(15,0.5293)(18,0.5371)};
\addplot[advtrain] coordinates {(0,0.5112)(3,0.5352)(6,0.5683)(9,0.5902)(12,0.5901)(15,0.5929)(18,0.6102)};
\addplot[semrect] coordinates {(0,0.4460)(3,0.4767)(6,0.5028)(9,0.5259)(12,0.5549)(15,0.5623)(18,0.5638)};
\addplot[signdeepsc] coordinates {(0,0.5622)(3,0.5855)(6,0.6067)(9,0.6213)(12,0.6336)(15,0.6426)(18,0.6521)};
\addplot[gray, dashed, very thick, mark=o, mark size=2pt, every mark/.append style={solid}] coordinates {(0,0.4547)(3,0.4664)(6,0.4787)(9,0.4924)(12,0.5023)(15,0.5076)(18,0.5063)};
\addplot[orange!80!black, dashed, very thick, mark=square, mark size=2pt, every mark/.append style={solid}] coordinates {(0,0.4838)(3,0.5133)(6,0.5266)(9,0.5370)(12,0.5422)(15,0.5540)(18,0.5544)};
\addplot[violet, dashed, very thick, mark=triangle, mark size=2.5pt, every mark/.append style={solid}] coordinates {(0,0.4274)(3,0.4491)(6,0.4776)(9,0.5000)(12,0.5173)(15,0.5321)(18,0.5325)};
\addplot[blue!70!black, dashed, very thick, mark=diamond, mark size=2.5pt, every mark/.append style={solid}] coordinates {(0,0.5557)(3,0.5818)(6,0.6034)(9,0.6145)(12,0.6462)(15,0.6503)(18,0.6588)};
\end{axis}
\end{tikzpicture}
\end{tabular}
\\[2pt]
\begin{tikzpicture}
\begin{axis}[
  hide axis, width=0.9\textwidth, height=2cm,
  xmin=0, xmax=1, ymin=0, ymax=1,
  legend columns=6,
  legend style={font=\scriptsize, draw=none, /tikz/every even column/.append style={column sep=8pt}, at={(0.5,0.5)}, anchor=center},
]
\addplot[gray, mark=*, mark size=2pt, very thick] coordinates {(0,0)};
\addplot[orange!80!black, mark=square*, mark size=2pt, very thick] coordinates {(0,0)};
\addplot[violet, mark=triangle*, mark size=2.5pt, very thick] coordinates {(0,0)};
\addplot[blue!70!black, mark=diamond*, mark size=2.5pt, very thick] coordinates {(0,0)};
\addplot[black, very thick] coordinates {(0,0)};
\addplot[black, dashed, very thick] coordinates {(0,0)};
\legend{DeepSC, AdvTrain, SemRECT, SignDeepSC, FGSM, PGD}
\end{axis}
\end{tikzpicture}
\caption{Rayleigh channel performance vs.\ SNR, $\epsilon=0.7$. Top: clean-channel; bottom: under attack (solid\,=\,FGSM, dashed\,=\,PGD).}
\label{fig:rayleigh_mc}
\end{figure*}
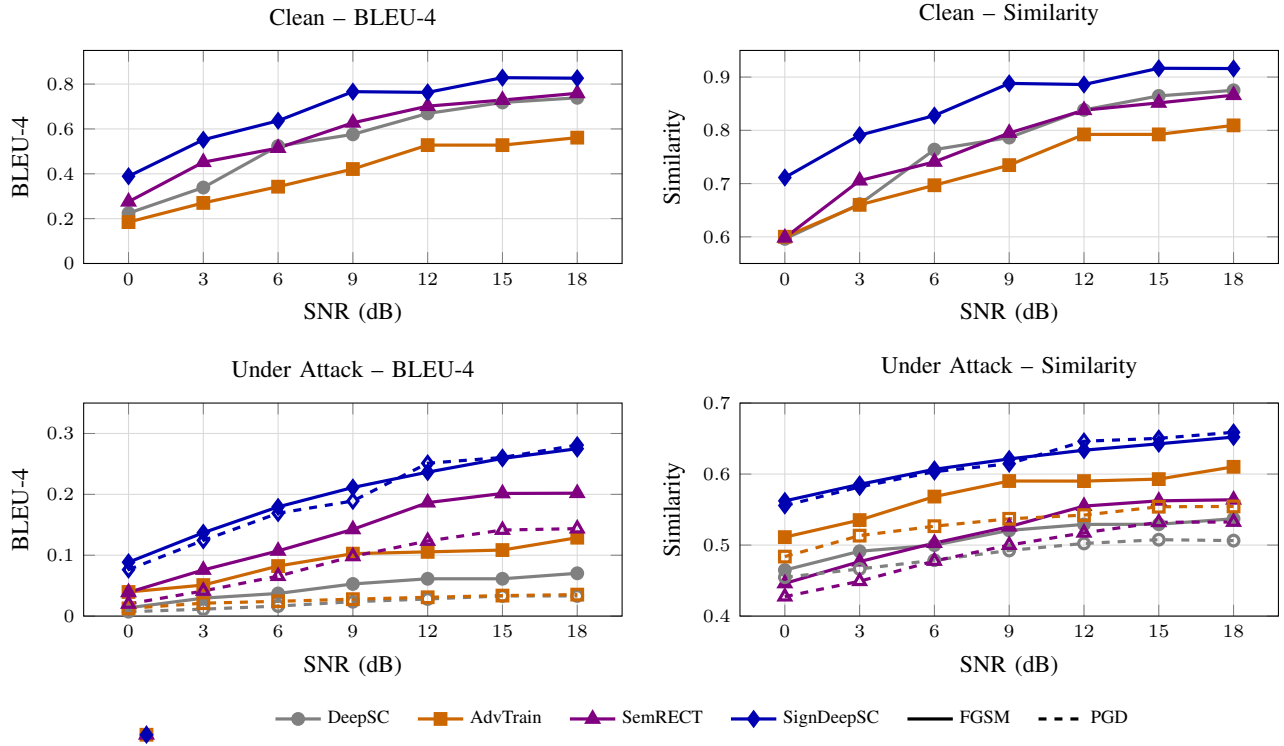

\textit{Comparison with baselines:}
Figure~\ref{fig:rayleigh_mc} shows performance across SNR for the Rayleigh channel. At SNR\,=\,12\,dB (3-seed Monte Carlo means), DeepSC collapses under attack (FGSM BLEU\,=\,0.058). AdvTrain recovers partially but incurs a 13\% clean penalty ($0.682{\to}0.591$), and PGD BLEU reaches only 0.084.
SemRECT achieves slightly higher FGSM BLEU (0.169 vs.\ 0.155) but lower similarity (0.541 vs.\ 0.624): its GAN recovers some n-grams (i.e., short word sequences) but fuses them poorly, while AdvTrain outputs more coherent sentences. SignDeepSC leads across all metrics without adversarial training (FGSM BLEU\,=\,$0.233{\pm}0.011$, Sim\,=\,$0.631{\pm}0.005$; PGD BLEU\,=\,$0.237{\pm}0.005$). Because the original encoder is frozen and the signature acts as additive side information, robustness improves without constraining the main feature path, resulting in no clean-channel penalty.

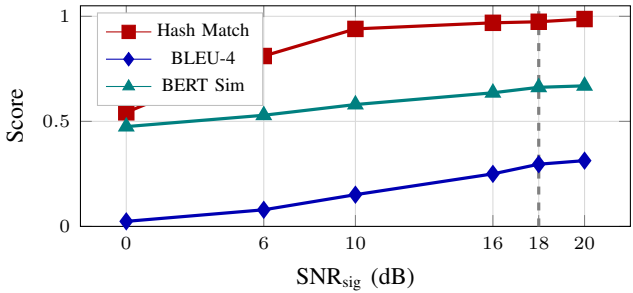
\begin{figure}[t]
\centering
\begin{tikzpicture}
\begin{axis}[
  width=\columnwidth, height=4.5cm,
  grid=major, grid style={gray!30},
  xlabel={$\text{SNR}_{\text{sig}}$ (dB)},
  ylabel={Score},
  xtick={0,6,10,16,18,20},
  ymin=0, ymax=1.05,
  tick label style={font=\scriptsize},
  label style={font=\small},
  legend style={font=\scriptsize, at={(0.03,0.97)}, anchor=north west, draw=gray!50, thin},
  axis line style={thin},
  very thick,
]
\addplot[red!70!black, mark=square*, mark size=2.5pt] coordinates {(0,0.543)(6,0.811)(10,0.940)(16,0.969)(18,0.974)(20,0.987)};
\addplot[blue!70!black, mark=diamond*, mark size=2.5pt] coordinates {(0,0.024)(6,0.079)(10,0.151)(16,0.250)(18,0.296)(20,0.313)};
\addplot[teal, mark=triangle*, mark size=2.5pt] coordinates {(0,0.475)(6,0.529)(10,0.580)(16,0.636)(18,0.662)(20,0.669)};
\draw[gray, dashed] (axis cs:18,0) -- (axis cs:18,1.05);
\legend{Hash Match, BLEU-4, BERT Sim}
\end{axis}
\end{tikzpicture}
\caption{Signature channel sensitivity at main SNR\,=\,12\,dB, Rayleigh fading. Dashed line: training point ($\text{SNR}_{\text{sig}}{=}18$\,dB).}
\label{fig:sensitivity}
\end{figure}

\textit{Signature channel sensitivity:}
Fig.~\ref{fig:sensitivity} sweeps $\text{SNR}_{\text{sig}}$ from 0 to 20\,dB at a main SNR of 12\,dB (Rayleigh, no adversarial perturbation). The hash match rises sharply from 54\% at 0\,dB to $\geq$94\% above 10\,dB; BLEU and similarity follow the same trend, confirming that a moderate-quality side channel suffices.

\begin{table}[t]
\centering
\caption{AWGN channel, SNR\,=\,12\,dB, $\epsilon=0.7$. All models trained on Rayleigh only. Mean over 3 seeds; std\,$<$\,0.003 for all entries.}
\label{tab:awgn}
\setlength{\tabcolsep}{1.5pt}
\small
\begin{tabular}{l cc cc cc}
\toprule
& \multicolumn{2}{c}{\textbf{Clean}} & \multicolumn{2}{c}{\textbf{FGSM}} & \multicolumn{2}{c}{\textbf{PGD}} \\
\cmidrule(lr){2-3} \cmidrule(lr){4-5} \cmidrule(lr){6-7}
\textbf{Model} & BLEU & Sim & BLEU & Sim & BLEU & Sim \\
\midrule
DeepSC     & 0.739 & 0.875 & 0.070 & 0.537 & 0.033 & 0.510 \\
AdvTrain   & 0.659 & 0.850 & 0.184 & 0.645 & 0.093 & 0.614 \\
SemRECT    & 0.764 & 0.869 & 0.200 & 0.562 & 0.145 & 0.532 \\
SignDeepSC & \textbf{0.833} & \textbf{0.918} & \textbf{0.278} & \textbf{0.654} & \textbf{0.278} & \textbf{0.659} \\
\bottomrule
\end{tabular}
\end{table}

\textit{Cross-channel generalization:}
All models are trained on Rayleigh fading only. Table~\ref{tab:awgn} reports performance on AWGN at SNR\,=\,12\,dB, $\epsilon=0.7$. SignDeepSC leads across all six metrics without any clean-accuracy penalty. All models improve relative to Rayleigh fading, confirming cross-channel generalization without retraining.

\begin{table}[t]
\centering
\caption{Component ablation at SNR\,=\,12\,dB, Rayleigh fading, $\epsilon{=}0.7$}
\label{tab:ablation}
\setlength{\tabcolsep}{1.5pt}
\small
\begin{tabular}{l cc cc cc}
\toprule
& \multicolumn{2}{c}{\textbf{Clean}} & \multicolumn{2}{c}{\textbf{FGSM}} & \multicolumn{2}{c}{\textbf{PGD}} \\
\cmidrule(lr){2-3} \cmidrule(lr){4-5} \cmidrule(lr){6-7}
\textbf{Configuration} & BLEU & Sim & BLEU & Sim & BLEU & Sim \\
\midrule
No Signature            & 0.004 & 0.456 & 0.000 & 0.414 & 0.000 & 0.412 \\
No Repair (Sig+Scr)     & 0.004 & 0.457 & 0.001 & 0.426 & 0.002 & 0.431 \\
No Scrambler (Sig+Rep)  & 0.795 & 0.901 & 0.250 & 0.641 & 0.226 & 0.634 \\
Full SignDeepSC          & \textbf{0.778} & \textbf{0.893} & \textbf{0.375} & \textbf{0.711} & \textbf{0.537} & \textbf{0.783} \\
\bottomrule
\end{tabular}
\end{table}

\textit{Component ablation}: In Table~\ref{tab:ablation}, ablated variants are obtained by removing the named component(s) from the trained SignDeepSC model, without retraining; this isolates each component's learned contribution rather than the capacity of an architecture trained without it. The memory-replacement curriculum ($p_{\text{drop}}$ annealed from 0.7) trains the decoder to rely heavily on the signature-conditioned repair path, so removing it post hoc, rather than retraining without it, collapses BLEU even absent an attack; removing the signature also breaks descrambling.

\textit{Signature channel attack analysis:}
\label{sec:sig_attack}
We now relax the side-channel assumption to quantify resilience under compromise. 
\begin{equation}
\max_{\boldsymbol{\delta}_m,\,\boldsymbol{\delta}_s}\;\mathcal{L}_{\text{sem}}\!\bigl(\mathbf{T}_x{+}\boldsymbol{\delta}_m,\;\mathbf{T}_s{+}\boldsymbol{\delta}_s\bigr) \;\;\text{s.t.}\;\; \|\boldsymbol{\delta}_m\|_\infty{\le}\epsilon,\;\|\boldsymbol{\delta}_s\|_\infty{\le}\epsilon_{\text{sig}},\label{eq:joint}
\end{equation}
where $\mathbf{T}_s$ is the transmitted signature signal. For FGSM, both perturbations are computed in a single gradient step; for PGD, both are iteratively refined following~\eqref{eq:pgd}. Table~\ref{tab:joint_attack} fixes $\epsilon{=}0.7$ on the main channel and sweeps $\epsilon_{\text{sig}}$.

FGSM on the signature reduces BLEU by only 7\%, confirming robustness to weak side-channel interference. A white-box PGD attack degrades performance more sharply (0.238${\to}$0.050), but it requires access to the signature channel itself; because the signature is low-bandwidth, that channel can be protected efficiently using standard measures such as hopping or encryption.

\begin{table}[t]
\centering
\caption{Joint attack: Main $\epsilon{=}0.7$, varying $\epsilon_{\text{sig}}$, Rayleigh fading at 12\,dB.}
\label{tab:joint_attack}
\setlength{\tabcolsep}{3pt}
\small
\begin{tabular}{r cc cc}
\toprule
& \multicolumn{2}{c}{\textbf{FGSM}} & \multicolumn{2}{c}{\textbf{PGD}} \\
\cmidrule(lr){2-3} \cmidrule(lr){4-5}
$\epsilon_{\text{sig}}$ & BLEU-4 & Sim & BLEU-4 & Sim \\
\midrule
0.0 & 0.244 & 0.638 & 0.238 & 0.639 \\
0.1 & 0.259 & 0.653 & 0.050 & 0.553 \\
0.3 & 0.244 & 0.641 & 0.048 & 0.549 \\
0.5 & 0.226 & 0.627 & 0.046 & 0.541 \\
0.7 & 0.228 & 0.626 & 0.042 & 0.530 \\
\bottomrule
\end{tabular}
\end{table}

\section{Conclusion}

We proposed SignDeepSC, which leverages a compact semantic signature and cross-attention conditioning to defend semantic communication against physical-layer MitM attacks. Unlike adversarial training or GAN-based reconstruction, SignDeepSC requires no adversarial examples during training and operates with a single forward pass at inference. Experiments on Rayleigh fading and AWGN channels demonstrate consistent gains over all baselines under both clean and adversarial conditions. These gains hold primarily when the low-bandwidth signature channel is well protected.


{\small
\bibliographystyle{IEEEbib}
\bibliography{references}
}

\end{document}